\documentclass[final]{cvpr}

\usepackage{times}
\usepackage{epsfig}
\usepackage{graphicx}
\usepackage{amsmath}
\usepackage{amssymb}

\usepackage{subfigure}
\usepackage{multirow}
\usepackage{multicol}
\usepackage{threeparttable}
\usepackage{booktabs}
\usepackage{verbatim}

\usepackage[pagebackref=true,breaklinks=true,colorlinks,bookmarks=false]{hyperref}

\newcommand{\tabincell}[2]{\begin{tabular}{@{}#1@{}}#2\end{tabular}}

\def\cvprPaperID{58} 
\def\confYear{CVPR 2021}
\begin{document}

\title{Boosting the Performance of Video Compression Artifact Reduction \\ with Reference Frame Proposals and Frequency Domain Information}
\author{
	Yi Xu\textsuperscript{1}\footnotemark[2],
	Minyi Zhao\textsuperscript{1}\footnotemark[2],
	Jing Liu\textsuperscript{2}\footnotemark[2],
	Xinjian Zhang\textsuperscript{1},
	Longwen Gao\textsuperscript{2},
	Shuigeng Zhou\textsuperscript{1}\thanks{Corresponding author.},
	Huyang Sun\textsuperscript{2}\\
	{\small \textsuperscript{1}Shanghai Key Lab of Intelligent Information Processing, and School of Computer Science, Fudan University, China}\\
	{\small \textsuperscript{2}Bilibili, China}\\
	{\tt\small \{yxu17,zhaomy20,zhangxj17,sgzhou\}@fudan.edu.cn}\\
	{\tt\small \{liujing04,gaolongwen,sunhuyang\}@bilibili.com}

}

\maketitle
\renewcommand{\thefootnote}{\fnsymbol{footnote}}
\footnotetext[2]{Y. Xu, M. Zhao, J. Liu are co-first authors of the paper.}

\thispagestyle{empty}

\begin{abstract}
   Many deep learning based video compression artifact removal algorithms have been proposed to recover high-quality videos from low-quality compressed videos. Recently, methods were proposed to mine spatiotemporal information via utilizing multiple neighboring frames as reference frames. However, these post-processing methods take advantage of adjacent frames directly, but neglect the information of the video itself, which can be exploited. In this paper, we propose an effective reference frame proposal strategy to boost the performance of the existing multi-frame approaches. Besides, we introduce a loss based on fast Fourier transformation~(FFT) to further improve the effectiveness of restoration. Experimental results show that our method achieves better fidelity and perceptual performance on MFQE 2.0 dataset than the state-of-the-art methods. And \textbf{our method won Track 1 and Track 2, and was ranked the 2nd in Track 3 of NTIRE 2021 Quality enhancement of heavily compressed videos Challenge}.

\end{abstract}


\section{Instruction}
To handle the problems of huge storage cost and limited bandwidth while storing and transmitting multimedia data,
lossy compression algorithms are commonly used to compress multimedia data (\emph{e.g.} images, audios and videos).
These irreversible compression algorithms often introduce compression artifacts that degrade the quality of experience (QoE), especially for videos. Accordingly, video compression artifact removal, which aims to reduce the introduced artifact and recover details for lossy compressed videos, becomes a hot topic in the multimedia field~\cite{guan2019mfqe, xu2019non, deng2020spatio}.

\begin{figure}[t]
	\centering
	\includegraphics[width=0.92\linewidth]{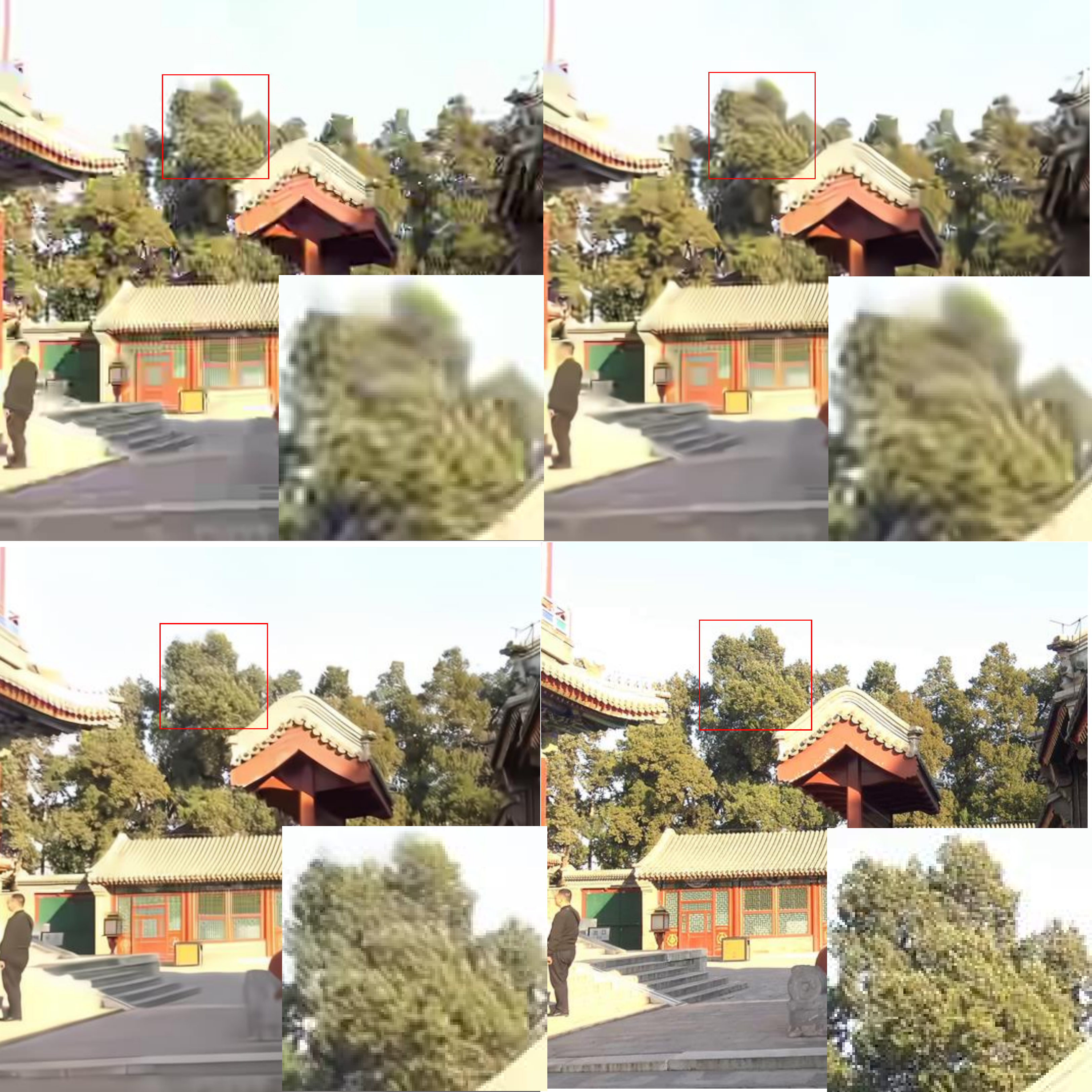}
	\caption{Examples of high-frequency recovery. These visual cases are compressed frame~(top left), prediction of STDF~(top right), prediction of a model trained with $L1$ + FFT loss~(\textbf{ours})~(bottom left), and the ground truth~(bottom right).}\label{fig-motivation}
\end{figure}

With the success of deep learning in text, image, and video processing, many deep neural network based compression artifact removal works have emerged and achieved significant performance improvement.
The rapid progress in this low-level task can be attributed to deep neural networks~\cite{dong2015compression, zhang2017beyond,li2017efficient,chen2018dpw,guo2016building}, various video compression priors~\cite{dai2017convolutional,jin2018quality,yang2017decoder}, and additional temporal information~\cite{yang2018multi,lu2018deep, yang2019quality, guan2019mfqe,xu2019non, lu2019deep,deng2020spatio}, respectively.
Among them, \cite{dong2015compression, li2017efficient, zhang2017beyond}  are  designed for JPEG compression artifact removal and can be adopted for videos by restoring each frame individually.
\cite{dai2017convolutional, yang2017decoder, jin2018quality} are proposed based on the fact that I/P/B frames are compressed with different strategies and should be restored by individual  models.  These methods utilize a single frame as input but neglect temporal dependency with  neighboring frames. To remedy this drawback, \cite{yang2018multi, guan2019mfqe} exploit two motion-compensated nearest peak-quality frames~(PQFs) as reference frames, \cite{lu2018deep, lu2019deep} develop the deep Kalman filter network and capture spatiotemporal information from preceding frames, and \cite{xu2019non, deng2020spatio} employ the non-local ConvLSTM and deformable convolution respectively, to capture dependency among multiple neighboring frames.
However, using only preceding frames omits the information from the followings;  restoration with a pair of nearby PQFs leads to the missing of high-quality details from some other frames~(as mentioned in \cite{xu2019non}). Recent methods~\cite{xu2019non, deng2020spatio} circumvent this problem but directly utilize the multiple adjacent frames as reference frames.

This paper is a summary of our method developed for NTIRE 2021 Quality enhancement of heavily compressed videos Challenge. We formulate an effective Reference Frame Proposal~(abbreviated as RFP) strategy as an incremental technique equipped for methods incorporating multiple frames in this task. It is natural for RFP to be applied to \cite{xu2019non, deng2020spatio}.
Considering that \cite{xu2019non} suffers severe computation and memory costs and is hard to be extended to very deep models used for the Challenge, we applied our RFP to another state-of-the-art method STDF~\cite{deng2020spatio} during the competition.  Besides, as shown in Fig.~\ref{fig-motivation}, over-smoothing harms the performance of enhanced frames a lot. Details and textures are almost removed after enhanced by STDF.
The over-smoothing phenomenon indicates that high-frequency details are dropped~\cite{chen2018dpw, cui2019decoder, luo2020deep, liu2020trident}, thus we introduce an additional optimization objective based on fast Fourier transformation (FFT) to supervise the learning of frequency domain information. That is, we exploit both spatial and frequency supervision signals to train the model and complement missing details. Empirical experiments show that both the RFP strategy and the FFT loss lead to significant performance improvement, and combining these two techniques can further boost the performance.
Moreover, we adopt a very deep Quality Enhancement~(QE) module based on \cite{yu2018wide, Fuoli_2020_CVPR_Workshops} in the competition.
In summary, the contributions of this work are as follows:
\begin{enumerate}
\item We propose an effective Reference Frame Proposal strategy by utilizing the neighboring compressed frames, which can be directly equipped for existing multi-frame approaches.

\item We introduce a loss based on FFT in this task to complement the missing high-frequency details.

\item We adopt an effective architecture for QE module, which can perform superior results with similar FLOPs and be extended to very deep models.
	
\item We conduct extensive experiments over MFQE 2.0 dataset and achieve state-of-the-art performance. Our solution is the winner of Track 1 and Track 2 - (Fixed QP, Fidelity /  Perceptual), and won the 2nd place in Track 3 - (Fixed bit-rate Fidelity) of \textbf{NTIRE 2021 Quality enhancement of heavily compressed videos Challenge}.
\end{enumerate}

\begin{figure*}
	\centering
	\includegraphics[width=0.9\linewidth]{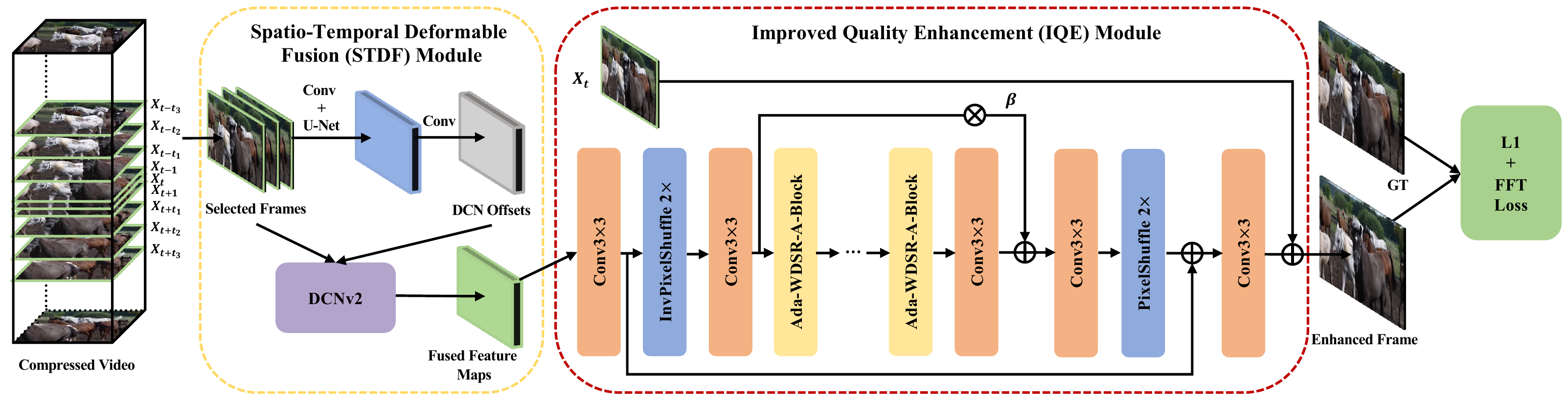}
	\caption{The framework of our method.}\label{fig-framework}
\end{figure*}

\section{Related Work}
In this section, we review the related work of compression artifact reduction based on deep-learning techniques.
Following the success of deep learning on ImageNet~\cite{russakovsky2015imagenet}, many methods with neural networks have been proposed in this long-standing low-level task. According to the utilization of domain knowledge and the number of input frames, existing methods can be categorized into three groups: image-based approaches, single-frame, and multi-frame approaches, respectively.

\textbf{Image-based Approaches}. These approaches are proposed for image compression artifact removal~\cite{dong2015compression, zhang2017beyond, li2017efficient, galteri2017deep,  yoo2018image, chen2018dpw, guo2017one, liu2018non, zhang2019residual}. When applied to the compressed videos, these methods are fed with a single frame and enhance it without knowledge of the video compression algorithms. For example, ARCNN~\cite{dong2015compression}  is the first work proposed for reducing JPEG compression artifacts. There are four convolutional layers without any pooling or fully connected layers.
DnCNN~\cite{zhang2017beyond} is another typical method that exploits deeper networks with batch normalization and residual learning. More recently, \cite{yoo2018image, chen2018dpw} enhance visual quality via wavelet/frequency domain information.
\cite{liu2018non, zhang2019residual} utilize the non-local mechanism for restoration in low-level tasks.

\textbf{Single-frame Approaches}. Some of such approaches~\cite{wang2017novel, dai2017convolutional, yang2017decoder, yang2018enhancing} employ knowledge of different coding modes in video compression algorithms, \emph{e.g.} I/P/B frames. However, these methods omit the temporal information in frame sequence, and they are ineffective in handling some kinds of temporal noise, such as mosquito noise, edge floating, and flickering. Specifically, DS-CNN~\cite{yang2017decoder} and QE-CNN~\cite{yang2018enhancing} were proposed with two independent models, and they are responsible for intra coding and inter coding modes, respectively.

\textbf{Multi-frame Approaches}.
\cite{lu2018deep, lu2019deep} model this vision task as a Kalman filtering procedure, enhancing the frame sequence recursively and capturing temporal information from enhanced preceding frames. \cite{lu2018deep, lu2019deep} further incorporate quantized prediction residual in compressed code streams as strong prior knowledge. However, exploiting temporal information from only preceding frames is incomplete because B frames are compressed via preceding and following frames. Given that the quality of compressed frames in videos fluctuates dramatically, \cite{yang2018multi, guan2019mfqe} proposed MFQE to build temporal dependency with nearby higher-quality frames. In the MFQE series methods, a classifier is first employed for detecting PQFs, then PQFs are enhanced without reference frames, while non-PQFs take these PQFs as reference frames, compensate reference frames with optical flow and utilize a slow-fused strategy to capture spatial and temporal information from PQFs. Later, \cite{yang2019quality} was proposed with a modified convolutional LSTM. Due to the limitation of motion flow and the observation that high-quality patch also exists in nearby low-quality frames, \cite{xu2019non, deng2020spatio} utilize the non-local mechanism or deformable convolutional network for capturing spatiotemporal dependency in multiple adjacent frames.

\textbf{Difference between Our Method and the Existing Multi-frame Ones}.
Mining spatiotemporal information from multiple frames becomes a trend for the quality enhancement of compressed videos. However, the state-of-the-art methods select reference frames in a naive form. In our method, a guidance technique is introduced for reference frame proposals in the preliminary step. Besides, to remedy the over-smoothing phenomenon in this task, an additional loss based on FFT is developed to help recover high-frequency details. Furthermore, we utilize a very deep model based on \cite{yu2018wide, Fuoli_2020_CVPR_Workshops} in the QE module.

\section{Method}
As for multi-frame approaches, most of them can be concluded as three essential components: \emph{Reference Frame Proposal} (RFP) module, \emph{Spatio-Temporal Feature Fusion} (STFF) module, and \emph{Quality Enhancement} (QE) module. Recently, multi-frame approaches focus on improving the STFF module but still employ a naive reference frame proposal strategy in the RFP module. Thus, in this paper, we pay more attention to the other modules and loss function.

\subsection{Reference Frame Proposal}
The goal of video compression artifact reduction is to produce a high-quality frame $\hat{Y}_t$ from a compressed frame $X_t$ of the original frame (the ground truth) $Y_t$, where $X_t \in \mathbb{R}^{C \times H\times W}$, $C$ is the number of channels of a single frame,
$H$ and $W$ are the width and height of input videos.
In the RFP module, we need to select $2R$ frames from the compressed sequence $X=\{X_1, X_2, \cdots, X_t, \cdots, X_T\}$ as reference frames $\{X_{t+{t}_1}, \cdots, X_{t+{t}_{2R}}\}$ for the target frame $X_t$. Here, the first $R$ frames $\{X_{t+t_1}, \cdots, X_{t+t_R}\}$ are the preceding frames of the target frame $X_t$, $\{X_{t+t_{R+1}}, \cdots, X_{t+t_{2R}}\}$ are the following frames, and $R$ is the number of reference frames in one direction. For the sake of simplicity, we take the preceding frames as example in the following.

Assume that $\{X_{t+t_1}, ..., X_{t+t_R}\}$ is an ordered sequence, and $t_1 < \cdots < t_R < 0$.  Then, the rules of RFP can be described as follows:

1) As the preliminary step, we first extract the metadata from the HEVC bit-stream with  HM-Decoder. The encoder configurations in Track 1/2 and Track 3 are different. Thus, we obtain the candidate frames for RFP with different metadata from bit-stream. In Track 1/2, we set the frame whose QP score is lower than that of the two adjacent frames as candidate frame. While in Track 3, all I/P frames are regarded as candidate frames.

2) We fixedly select adjacent frame $X_{t-1}$ of $X_t$ as the first reference frame by setting $t_R=-1$.

3) We  recursively take the next preceding candidate frame of the last selected reference frame as a new reference frame until there are $R$ reference frames or no candidate frames are left.

4) If there is no more candidate frame and the number of selected reference frames is smaller than $R$, then repeatedly pad it with the last selected frame until there are $R$ frames.

\begin{figure}
	\centering
	\includegraphics[width=0.9\linewidth]{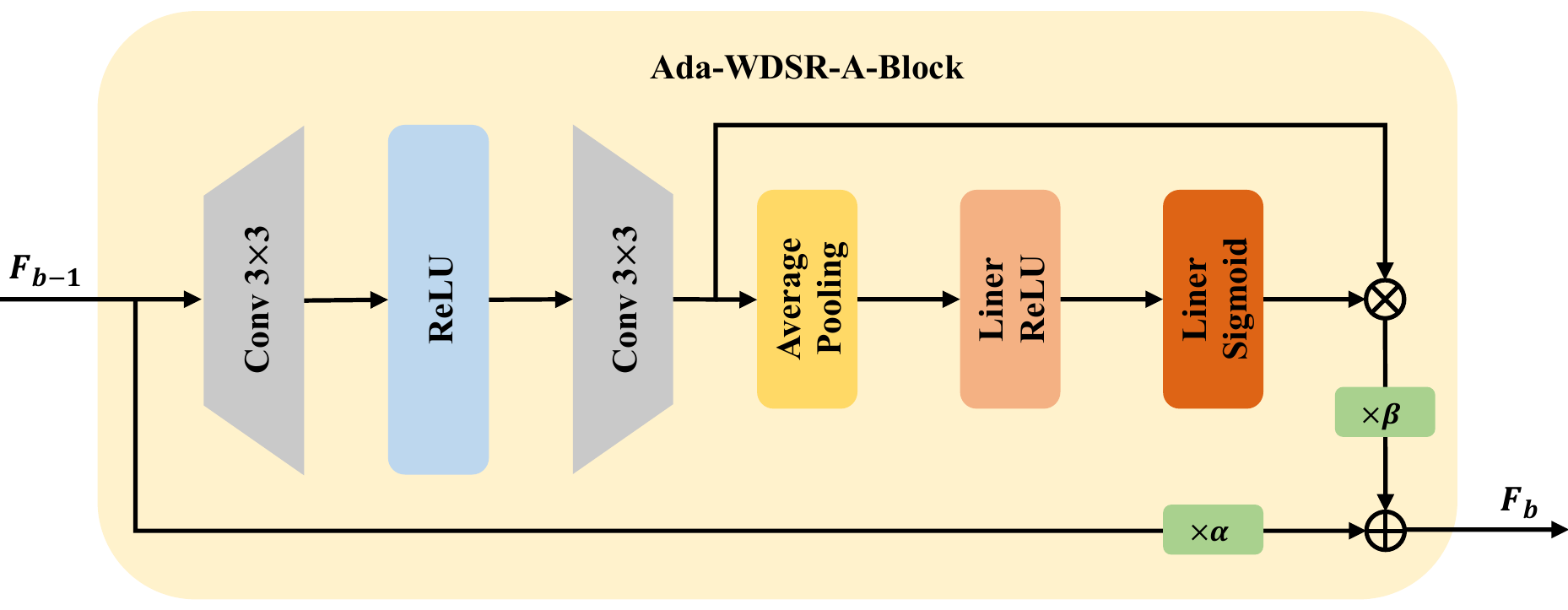}
	\caption{The architecture of Ada-WDSR-A-Block.}
	\label{fig-block}
\end{figure}

\subsection{Improved Quality Enhancement Module}
\label{sec-IQE}
The gist of the QE module is to take the fused feature from the STFF module~(\emph{i.e.}, the STDF module in Fig.~\ref{fig-framework}) as input and produce a residual, which is used together with the compressed frame to reconstruct the enhanced frame. Apart from the STFF module, the QE module is another critical factor for artifact reduction because it needs to explore spatiotemporal information and complement introduced artifacts.
For a fair comparison, we employ the same QE module on the benchmark dataset MFQE 2.0. Furthermore, we adopt an Improved Quality Enhancement~(IQE) module, which is extended to a very deep version to achieve better results in the challenge.

The framework of the IQE module is shown in Fig.~\ref{fig-framework}. First, we utilize a global residual connection between the head and tail convolutional layers. Parallel to the global connection, the architecture consists of a down-sample module, a deep backbone, and an up-sample module. Among them, the down-sample module is composed of an inverse pixel shuffle layer~\cite{shi2016real} and a convolutional layer to decrease the spatial resolution, and the up-sample module utilizes an architecture contrast to that of the down-sample module. Between them, there is a skip connection with fixed residual scale $\beta=0.2$ and a stack of Adaptive WDSR-A-Blocks~\cite{yu2018wide}~(Ada-WDSR-A-Block) followed by a convolutional layer.

The Ada-WDSR-A-Blocks are utilized to explore the complementary information for the compressed frame in our paper. The structure of Ada-WDSR-A-Block is illustrated in Fig.~\ref{fig-block}. For all Ada-WDSR-Blocks,  scale $r$ is set to $4$. Comparing with WDSR-A-Blocks, there are two additional learnable parameters $\alpha$ and $\beta$ in Ada-WDSR-Block, which are initialized with 1 and 0.2, respectively. Additionally, we deploy a channel-attention layer~\cite{zhang2018image} before re-scaling the body stream in Ada-WDSR-A-Blocks.

\subsection{Fast Fourier Transformation loss}
To complement the missing high-frequency details caused by over-smoothing, we introduce a novel supervision signal based on fast Fourier transformation as a complementary loss. Concretely, we apply the fast Fourier transformation to both the ground truth $Y_t$ and the prediction of the QE module, and then employ $L1$ loss on both amplitude and phase of them. The amplitude $A(\cdot)$ and the phase $P(\cdot)$ of a given frame $X$ are calculated as follows:
\begin{equation}
	\label{eq-fft}
	\begin{split}
		X^f(u,v) &= \sum_x^H{\sum_y^W{X(x, y)e^{-\emph{j} \ 2\pi (\frac{u * x}{H} + \frac{v * y}{W})}}},\\	
		A(X) &= \sqrt{Re(X^f)^2 + Im(X^f)^2}, \\
		P(X) &= atan(Im(X^f) / Re(X^f)), \\
	\end{split}	
\end{equation}
where $X(x, y)$ denotes the value at spatial position $(x, y)$,  $Re(\cdot)$ and $Im(\cdot)$ are the real and imaginary parts of $X^f$.  Accordingly, we have the following loss function:
\begin{equation}
	\mathcal{L}_{FFT} = \| A(\hat{Y}) - A(Y) \|_2 + \lambda \| P(\hat{Y}) - P(Y) \|_2, \\
	\label{eq-lossfft}
\end{equation}
where $\lambda$ is a trade-off hyper-parameter between amplitude and phase ($\lambda=1$ in our implementation).
With the FFT loss as a complementary supervision signal, our model is more powerful in high-frequency detail recovery.

\section{Experiments}
Actually, our techniques can be used in most multi-frame approaches.
Here, we take the state-of-the-art STDF~\cite{deng2020spatio} as an example for evaluating our techniques.
We conduct extensive experiments on the MFQE 2.0 dataset and the dataset provided by the competition.
Our evaluation consists of three parts: 1) Ablation study on NTIRE 2021 Dataset~\cite{yang2021ntire_dataset}; 2) Comparison with state-of-the-art methods on MFQE 2.0 dataset~\cite{guan2019mfqe} with five QPs;
3) Performance of our method on three tracks in NTIRE 2021~\cite{yang2021ntire_method}.

\subsection{Datasets and Settings}
\label{sec-dataset}

\textbf{MFQE 2.0 dataset~\cite{guan2019mfqe}.} It consists of 126 video sequences collected from Xiph.org~\cite{Xiph}, VQEG~\cite{VQEG} and JCT-VC~\cite{bossen2011common}. Resolutions of these video sequences vary from $352\times 240$ to $2560\times 1680$. For a fair comparison, we follow the settings in \cite{guan2019mfqe, deng2020spatio}: 108 of them are taken for training and the remaining 18 for testing. All sequences are encoded in HEVC Low-Delay-P (LDP) configuration, using HM 16.20 with $QP$=22, 27, 32, 37 and 42.

\textbf{NTIRE 2021 Dataset~\cite{yang2021ntire_dataset}.} It is provided in NTIRE 2021 Quality enhancement of heavily compressed videos Challenge~\cite{yang2021ntire_method}. There are 200 videos for training in the competition, 20 for validation, and 20 for the final test. However, only videos in the training data are provided with uncompressed videos. Thus, in this paper, we split 200 videos into two parts: 190 for training and 10 for validation. Sequences are encoded in HEVC LDP configuration with $QP$=37 in Track 1 and 2, and encoded by FFmpeg supported with libx265 with fixed bit-rate 200kbps in Track 3. Our ablation study in Sec.~\ref{sec-abstudy} is evaluated with the settings in Track 1.

\subsection{Implementation Details}
In this paper, we take the state-of-the-art method STDF~\cite{deng2020spatio} as our baseline and conduct experiments by following the scheme of STDF.
To achieve similar FLOPs of the IQE module to that of the QE module~(R3L in STDF), we implement a shallow IQE module with 30 Ada-WDSR-A-Blocks, features in Ada-WDSR-Blocks are implemented with \{32, 128, 32\} channels in Sec.~\ref{sec-abstudy} and Sec.~\ref{sec-sota}.
For all datasets, models are trained by the Adam optimizer with an initial learning rate of $10^{-4}$, which is decreased by half when 60\% and 90\% iterations are finished.

\begin{table}
	\caption{Ablation study of our method at $QP$=$37$ over STDF. Experiments are presented with $\Delta$PSNR (dB) and $\Delta$SSIM ($\times 10^{-2}$) on validation sequences from NTIRE2021.}
	\centering
	\begin{tabular}{cccc|c}
		\toprule
		\tabincell{c}{RFP} & \tabincell{c}{L1} & \tabincell{c}{L2}  & \tabincell{c}{FFT} & \tabincell{c}{PSNR / SSIM} \cr
		\midrule
		- & - & $\checkmark$ & -  &  0.72 / 1.572 \\
		$\checkmark$ & - & $\checkmark$ & - & 0.74 / 1.607 \\
		- & $\checkmark$ & - & - & 0.68 / 1.497 \\
		- & - & $\checkmark$ & $\checkmark$  & 0.74 / 1.581 \\
		- & $\checkmark$ & - & $\checkmark$  & 0.74 / 1.610 \\
		$\checkmark$ & $\checkmark$ & - & $\checkmark$ &  \textbf{0.76} / \textbf{1.639} \\
		\bottomrule
	\end{tabular}
	\label{tab-abstudy}
	
\end{table}

\begin{figure}
	\centering
	\subfigure[b][Visual case of RFP]{
		\centering
		\includegraphics[width=0.92\linewidth]{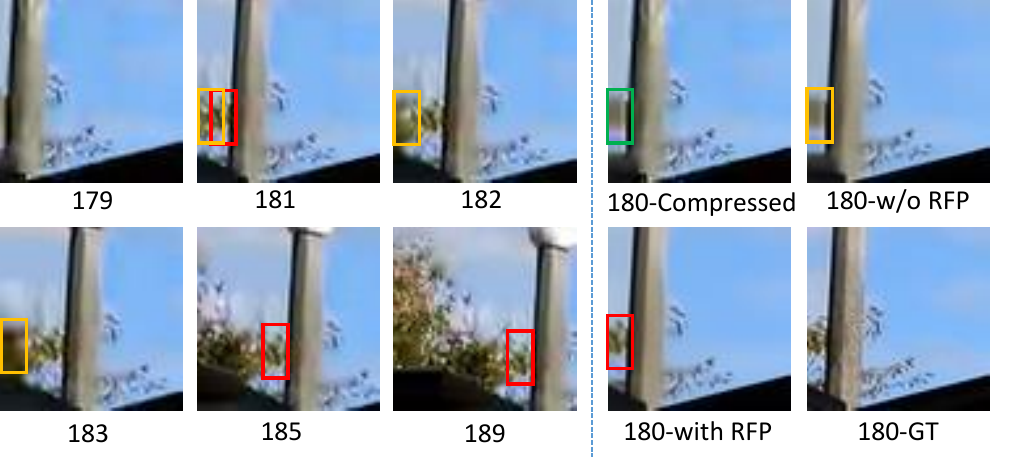}
		\label{fig-abstudy-a}
	}  \\
	\subfigure[b][Visual case of FFT loss]{
		\centering
		\includegraphics[width=0.9\linewidth]{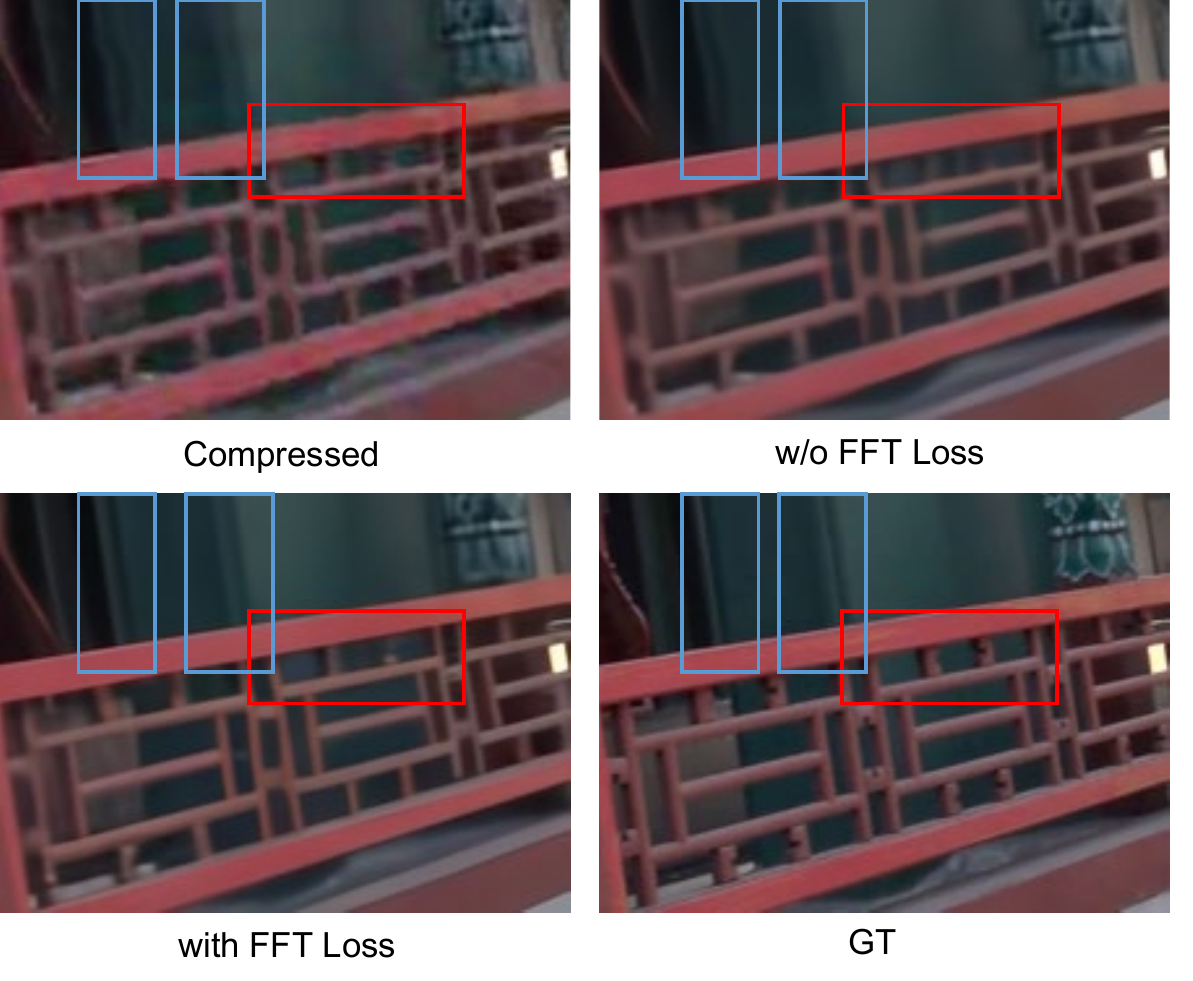}
		\label{fig-abstudy-b}
	}
	\caption{Visual examples of ablation study. a) Results of RFP. Frames with the yellow box are the reference frames used by the original STDF. Frames with the red box are the references proposed by RFP. Reference frames from RFP provide additional details of the tree to enhance the compressed frame. b) Results of the FFT loss. Improvement by the FFT loss in visual perspective is bounded with rectangles.}
	\label{fig-abstudy}
\end{figure}

\subsection{Ablation Study}
As mentioned in Sec.~\ref{sec-dataset}, experiments of ablation study in this paper are conducted on the dataset from NTIRE 2021 with settings in Track 1. Experimental results included in Tab.~\ref{tab-abstudy} are STDF and STDF with our techniques.
Among them, results with $L2$ loss in the second row of Tab.~\ref{tab-abstudy} is the performance of baseline STDF. As listed in Tab.~\ref{tab-abstudy}, all experiments except that for the loss function and RFP strategy
follow the same setting as STDF.

\label{sec-abstudy}
\textbf{Effect of reference frame proposal.} Here, we evaluate the effectiveness of utilizing our RFP strategy. First, we compare the performance between STDF (using $L2$ loss) and STDF with our RFP strategy (using RFP + $L2$ loss). The results in the 2nd and 3rd row in Tab.~\ref{tab-abstudy} show that utilization of RFP in STDF can improve the performance effectively. Visual examples in Fig.~\ref{fig-abstudy-a} also show that utilizing RFP brings benefit by learning missing details from adjacent $2R$ frames. Then, we further verify the effectiveness of RFP on the model trained with $L1$ and FFT loss. As shown in the last row in Tab.~\ref{tab-abstudy}, PSNR/SSIM achieves improvement over that with $L1$ and FFT loss (in the 6th row). A similar conclusion can also be obtained on the MFQE 2.0 dataset.
Thus, the results indicate that utilizing RFP makes the model achieve superior performance of restoration.

\textbf{Effect of FFT loss.}  Considering that $L1$ loss achieves better performance than $L2$ loss in recently proposed low-level methods (\emph{e.g.} \cite{yu2018wide, zhang2018image}) for the super-resolution task, we investigate the combination of $L1$/$L2$ loss and FFT loss to evaluate the effectiveness of FFT loss. Different from the conclusion in \cite{yu2018wide, zhang2018image}, models trained with $L2$ loss achieve better performance than $L1$ loss in the 2nd and 4th rows in Tab.~\ref{tab-abstudy}.  However, the combination of $L1$ loss and FFT loss (in the 6th row) achieves a better result than the combination of $L2$ and FFT loss (in the 5th row).
Besides the example illustrated in Fig.~\ref{fig-motivation}, we present additional visual examples in Fig.~\ref{fig-abstudy-b} for a further validation on the FFT loss.

\begin{table*}[t]
	\renewcommand\arraystretch{1.3}
	\centering
	\scriptsize
	\caption{Overall comparison for $\Delta$PSNR (dB) and $\Delta$SSIM ($\times10^{-2}$) over test sequences at five QPs.}
	\begin{threeparttable}
		
		\begin{tabular}{|c|c|l||c@{\hspace{0.4em}} c@{\hspace{0.4em}} c@{\hspace{0.4em}} c@{\hspace{0.4em}} c@{\hspace{0.4em}} c@{\hspace{0.4em}} c @{\hspace{0.4em}} c @{\hspace{0.4em}} | c |@{\hspace{0.4em}} c |}
			
			\hline
			\multirow{2}{*}{QP} & \multicolumn{2}{c||}{\multirow{2}{*}{Approach}} & AR-CNN & DnCNN &Li \textit{et al.} &DCAD & DS-CNN & {MFQE 1.0} & {MFQE 2.0} & {STDF}& Ours & Ours\\ [-0.3em]
			
			& \multicolumn{2}{c||}{} & {\cite{dong2015compression}}  & {\cite{zhang2017beyond}} & {\cite{li2017efficient}} & {\cite{wang2017novel}} & {\cite{yang2018enhancing}} & {\cite{yang2018multi}}  & {\cite{guan2019mfqe}}  & \cite{deng2020spatio} & \emph{same QE} & \emph{IQE} \\
			
			\hline
			\multirow{20}{*}{37} & \multicolumn{2}{c||}{Metrics} & \tiny PSNR / \tiny SSIM & \tiny PSNR / \tiny SSIM & \tiny PSNR / \tiny SSIM & \tiny PSNR / \tiny SSIM & \tiny PSNR / \tiny SSIM & \tiny PSNR / \tiny SSIM & \tiny PSNR / \tiny SSIM & \tiny PSNR / \tiny SSIM  & \tiny PSNR / \tiny SSIM  & \tiny PSNR / \tiny SSIM \\
			
			\cline{2-13}
			& \multirow{2}{*}{A} & \textit{Traffic}
			& 0.24 / 0.47 & 0.24 / 0.57 & 0.29 / 0.60 & 0.31 / 0.67 & 0.29 / 0.60 & 0.50 / 0.90 & 0.59 / 1.02 & 0.73 / 1.15 & 0.71 / \textbf{1.18} & \textbf{0.96} / \textbf{1.50} \\
			
			& & \textit{PeopleOnStreet}
			& 0.35 / 0.75 & 0.41 / 0.82 & 0.48 / 0.92 & 0.50 / 0.95 & 0.42 / 0.85 & 0.80 / 1.37 & {0.92} / {1.57} & 1.25 / 1.96 &\textbf{1.30} / \textbf{2.09} & \textbf{1.60} / \textbf{2.42} \\
			
			\cline{2-13}
			& \multirow{5}{*}{B} & \textit{Kimono}
			& 0.22 / 0.65 & 0.24 / 0.75 & 0.28 / 0.78 & 0.28 / 0.78 & 0.25 / 0.75 & 0.50 / 1.13 & {0.55} / {1.18} & 0.85 / 1.61 & \textbf{0.98} / \textbf{1.85} &  \textbf{1.09} / \textbf{2.01}\\
			
			& & \textit{ParkScene}
			& 0.14 / 0.38 & 0.14 / 0.50 & 0.15 / 0.48 & 0.16 / 0.50 & 0.15 / 0.50 & 0.39 / 1.03 & {0.46} / {1.23} & 0.59 / 1.47 & \textbf{0.62} / \textbf{1.58} & \textbf{0.79} / \textbf{2.00} \\
			
			& & \textit{Cactus}
			& 0.19 / 0.38 & 0.20 / 0.48 & 0.23 / 0.58 & 0.26 / 0.58 & 0.24 / 0.58 & 0.44 / 0.88 & {0.50} / {1.00} & 0.77 / 1.38 & 0.76 / \textbf{1.44} & \textbf{0.79} / \textbf{1.64} \\
			
			& & \textit{BQTerrace}
			& 0.20 / 0.28 & 0.20 / 0.38 & 0.25 / 0.48 & 0.28 / 0.50 & 0.26 / 0.48 & 0.27 / 0.48 & {0.40} / {0.67} & 0.63 / 1.06 & \textbf{0.65} / \textbf{1.08} & \textbf{0.67} / \textbf{1.16} \\
			
			& & \textit{BasketballDrive}
			& 0.23 / 0.55 & 0.25 / 0.58 & 0.30 / 0.68 & 0.31 / 0.68 & 0.28 / 0.65 & 0.41 / 0.80 & {0.47} / {0.83} & 0.75 / 1.23 & \textbf{0.86} / \textbf{1.43} & \textbf{0.91} / \textbf{1.90} \\
			
			\cline{2-13}
			& \multirow{4}{*}{C} & \textit{RaceHorses}
			& 0.22 / 0.43 & 0.25 / 0.65 & 0.28 / 0.65 & 0.28 / 0.65 & 0.27 / 0.63 & 0.34 / 0.55 & {0.39} / {0.80} & 0.55 / 1.35 & 0.55 / 1.34  & \textbf{0.58} / \textbf{1.61} \\
			
			& & \textit{BQMall}
			& 0.28 / 0.68 & 0.28 / 0.68 & 0.33 / 0.88 & 0.34 / 0.88 & 0.33 / 0.80 & 0.51 / 1.03 & {0.62} / {1.20} & 0.99 / 1.80 & \textbf{1.08} / \textbf{2.00} & \textbf{1.25} / \textbf{2.26} \\
			
			& & \textit{PartyScene}
			& 0.11 / 0.38 & 0.13 / 0.48 & 0.13 / 0.45 & 0.16 / 0.48 & 0.17 / 0.58 & 0.22 / 0.73 &{0.36} / {1.18} & 0.68 / 1.94 & 0.67 / 1.91 & \textbf{0.83} / \textbf{2.37} \\
			
			& & \textit{BasketballDrill}
			& 0.25 / 0.58 & 0.33 / 0.68 & 0.38 / 0.88 & 0.39 / 0.78 & 0.35 / 0.68 & 0.48 / 0.90 &{0.58} / {1.20} & 0.79 / 1.49 & \textbf{0.82} / \textbf{1.51} & \textbf{0.91} / \textbf{1.90}\\
			
			\cline{2-13}
			& \multirow{4}{*}{D} & \textit{RaceHorses}
			& 0.27 / 0.55 & 0.31 / 0.73 & 0.33 / 0.83 & 0.34 / 0.83 & 0.32 / 0.75 & 0.51 / 1.13 & {0.59} / {1.43} & 0.83 / 2.08 & \textbf{0.86} / \textbf{2.15} & \textbf{0.95} / \textbf{2.42} \\
			
			& & \textit{BQSquare}
			& 0.08 / 0.08 & 0.13 / 0.18 & 0.09 / 0.25 & 0.20 / 0.38 & 0.20 / 0.38 & -0.01 / 0.15 & {0.34} / {0.65} & 0.94 / 1.25 & 0.72 / 1.03 & \textbf{1.28} / \textbf{1.86}\\
			
			& & \textit{BlowingBubbles}
			& 0.16 / 0.35 & 0.18 / 0.58 & 0.21 / 0.68 & 0.22 / 0.65 & 0.23 / 0.68 & 0.39 / 1.20 & {0.53} / {1.70} & 0.74 / 2.26 & 0.72 / 2.21 & \textbf{0.91} / \textbf{2.88} \\
			
			& & \textit{BasketballPass}
			& 0.26 / 0.58 & 0.31 / 0.75 & 0.34 / 0.85 & 0.35 / 0.85 & 0.34 / 0.78 & 0.63 / 1.38 & {0.73} / {1.55} & 1.08 / 2.12 & \textbf{1.12} / \textbf{2.23} & \textbf{1.29} / \textbf{2.65} \\
			
			\cline{2-13}
			& \multirow{3}{*}{E} & \textit{FourPeople}
			& 0.37 / 0.50 & 0.39 / 0.60 & 0.45 / 0.70 & 0.51 / 0.78 & 0.46 / 0.70 & 0.66 / 0.85 & {0.73} / {0.95} & 0.94 / 1.17  & \textbf{1.00} / \textbf{1.28} & \textbf{1.24} / \textbf{1.50} \\
			
			& & \textit{Johnny}
			& 0.25 / 0.10 & 0.32 / 0.40 & 0.40 / 0.60 & 0.41 / 0.50 & 0.38 / 0.40 & 0.55 / 0.55 & {0.60} / {0.68} & 0.81 / 0.88 & \textbf{0.84} / \textbf{0.96} & \textbf{1.02} / \textbf{1.15}\\
			
			& & \textit{KristenAndSara}
			& 0.41 / 0.50 & 0.42 / 0.60 & 0.49 / 0.68 & 0.52 / 0.70 & 0.48 / 0.60 & 0.66 / 0.75 & {0.75} / {0.85} & 0.97 / 0.96 & \textbf{1.03} / \textbf{1.09} &  \textbf{1.23} / \textbf{1.23} \\
			
			\cline{2-13}
			& \multicolumn{2}{c||}{Average}
			& 0.23 / 0.45 & 0.26 / 0.58 & 0.30 / 0.66 & 0.32 / 0.67 & 0.30 / 0.63 & 0.46 / 0.88 & {0.56} / {1.09} & 0.83 / 1.51 & \textbf{0.85} / \textbf{1.58} & \textbf{1.03} / \textbf{1.90}\\
			
			\hline
			\hline
			42 & \multicolumn{2}{c||}{Average}
			& 0.29 / 0.96 & 0.22 / 0.77 & 0.32 / 1.05 & 0.32 / 1.09 & 0.31 / 1.01 & 0.44 / 1.30 & {0.59} / {1.65} & -- / -- & \textbf{0.79} / \textbf{2.18} &  \textbf{0.89} / \textbf{2.41}\\
			
			\hline
			32 & \multicolumn{2}{c||}{Average}
			& 0.18 / 0.19 & 0.26 / 0.35 & 0.28 / 0.37 & 0.32 / 0.44 & 0.27 / 0.38 & 0.43 / 0.58 & {0.516} / {0.68} & 0.86 / 1.04 & \textbf{0.93} / \textbf{1.16} & \textbf{1.08} / \textbf{1.36}\\
			
			\hline
			27 & \multicolumn{2}{c||}{Average}
			& 0.18 / 0.14 & 0.27 / 0.24 & 0.30 / 0.28 & 0.32 / 0.30 & 0.27 / 0.23 & 0.40 / 0.34 & {0.49} / {0.42} & 0.72 / 0.57  & \textbf{0.92} / \textbf{0.77} &  \textbf{1.09} / \textbf{0.92}\\
			
			\hline
			22 & \multicolumn{2}{c||}{Average}
			& 0.14 / 0.08 & 0.29 / 0.18 & 0.30 / 0.19 & 0.31 / 0.19 & 0.25 / 0.15 & 0.31 / 0.19 & {0.46} / {0.27} & 0.63 / 0.34 & \textbf{0.83} / \textbf{0.46} &  \textbf{0.96} / \textbf{0.53}\\
			\hline
			
		\end{tabular}

		\begin{tablenotes}
			\footnotesize
			\item[*] Video resolution: Class A ($2560\times1600$), Class B ($1920\times1080$), Class C ($832\times480$), Class D ($480\times240$), Class E ($1280\times720$).
		\end{tablenotes}
	\end{threeparttable}
\label{tab-sota-results}
\end{table*}

\subsection{Comparison with State-of-the-art Methods}
\label{sec-sota}
To demonstrate the advantage of our method, we compare the performance of our method and state-of-the-art approaches, including image-based~\cite{dong2015compression, zhang2017beyond, li2017efficient}, singe-frame~\cite{wang2017novel, yang2018enhancing} and multi-frame approaches~\cite{yang2018multi, guan2019mfqe, deng2020spatio}. For a fair comparison, our model is trained by following the training scheme of STDF.
 Results of video quality enhancement methods are cited from \cite{guan2019mfqe, deng2020spatio}. 

\textbf{Overall enhancement.}
Results of PSNR / SSIM improvement are presented in Tab.~\ref{tab-sota-results}. Here,  \emph{same QE} in Tab.~\ref{tab-sota-results} indicates the model follows the same architecture of the QE module in STDF, which means that the differences between \emph{same QE} and STDF are the REP strategy and the FFT loss. The improvement of \emph{same QE} over STDF can be regarded as the benefit from the REP strategy and the FFT loss. Meanwhile, the variant \emph{IQE} denotes the model with the improved QE module introduced in Sec.~\ref{sec-IQE}, from which a deeper version is designed as the final architecture used by us in the competition.

From Tab.~\ref{tab-sota-results}, we can see that all multi-frame approaches outperform the methods for images or single frames due to the benefit of utilizing spatiotemporal information. Moreover, the fact that STDF with an effective RFP strategy and FFT loss achieves better results than all the existing methods further proves the importance of filtering input information and the limitation of the $L2$ loss function. Moreover, \emph{IQE} further improves the performance on the benchmark dataset and achieves impressive results of $1.029dB / 0.0190$ PSNR/SSIM improvement for $QP$=37, 23.9\% and 25.8\% higher than that of STDF. Similar improvements can also be observed for other QPs.

\begin{figure}
	\centering
	\includegraphics[width=0.92\linewidth]{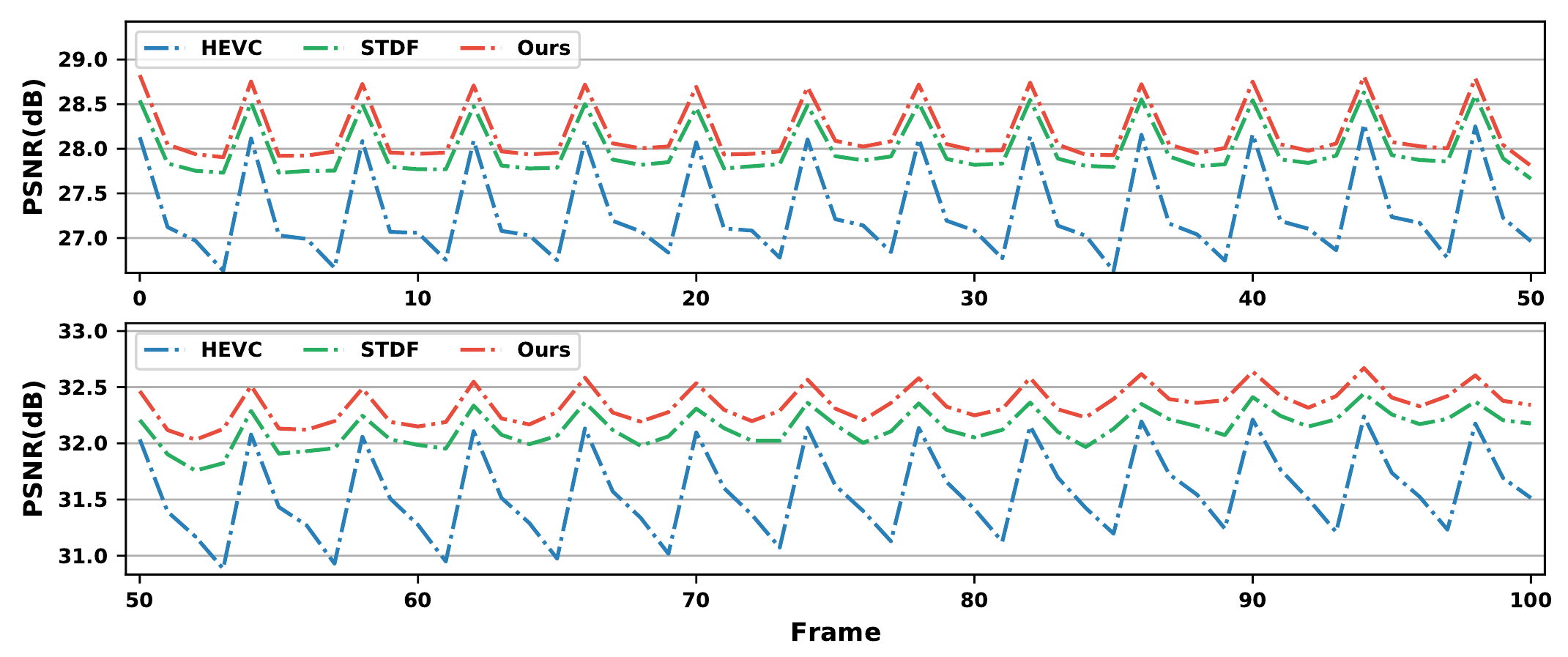}
	
	\caption{PSNR curves of HEVC baseline, STDF and ours on two test sequences at $QP$=37. Top: \emph{ParkScene}. Bottom: \emph{PartyScene}.}
	\label{fig-fluc}
\end{figure}

\begin{table}
	\caption{Average PVD/SD of test sequences for PSNR at $QP$=27, 32, 37 and 42.}
	\centering
	\scriptsize
	\begin{tabular}{c|cccc}
		\toprule
		\tabincell{c}{Method} & \tabincell{c}{27} & \tabincell{c}{32} & \tabincell{c}{37} & \tabincell{c}{42} \cr
		\midrule
		HEVC & 1.07 / 0.83 & 1.38 / 0.82 & 1.42 / 0.79 & 1.21 / 0.74 \\
		AR-CNN~\cite{dong2015compression} & 1.07 / 0.83 & 1.38 / 0.82 & 1.44 / 0.80 & 1.24 / 0.75 \\
		DnCNN~\cite{zhang2017beyond} & 1.06 / 0.83 & 1.40 / 0.83 & 1.44 / 0.80 & 1.24 / 0.75 \\
		Li \emph{et al.}~\cite{li2017efficient} & 1.06 / 0.83 & 1.38 / 0.83 & 1.44 / 0.80 & 1.24 / 0.76 \\
		DCAD~\cite{wang2017novel} & 1.07 / 0.83 & 1.39 / 0.83 & 1.45 / 0.80 & 1.26 / 0.76 \\
		DS-CNN~\cite{yang2018enhancing} & 1.07 / 0.83 & 1.39 / 0.83 & 1.46  / 0.80 & 1.24 / 0.75 \\
		MFQE 1.0~\cite{yang2018multi} & 0.84 / 0.81 & 1.07 / 0.77 & 1.05 / 0.73 & 0.82 / 0.69 \\
		MFQE 2.0~\cite{guan2019mfqe}& 0.77 / 0.74 & 0.98 / 0.70 & 0.96 / 0.67 & 0.74 / 0.62 \\
		Ours \emph{same QE} & \textbf{0.60} / \textbf{0.33} & \textbf{0.75} / \textbf{0.44} & \textbf{0.73} / \textbf{0.37} & \textbf{0.67} / \textbf{0.36}  \\
		ours \emph{IQE} & \textbf{0.58} / \textbf{0.32} & \textbf{0.72} / \textbf{0.47} & \textbf{0.70} / \textbf{0.41} & \textbf{0.66} / \textbf{0.30} \\
		\bottomrule
	\end{tabular}
	\label{tab-SDPVD}
	
\end{table}

\begin{figure*}
	\centering
	\includegraphics[width=0.92\linewidth]{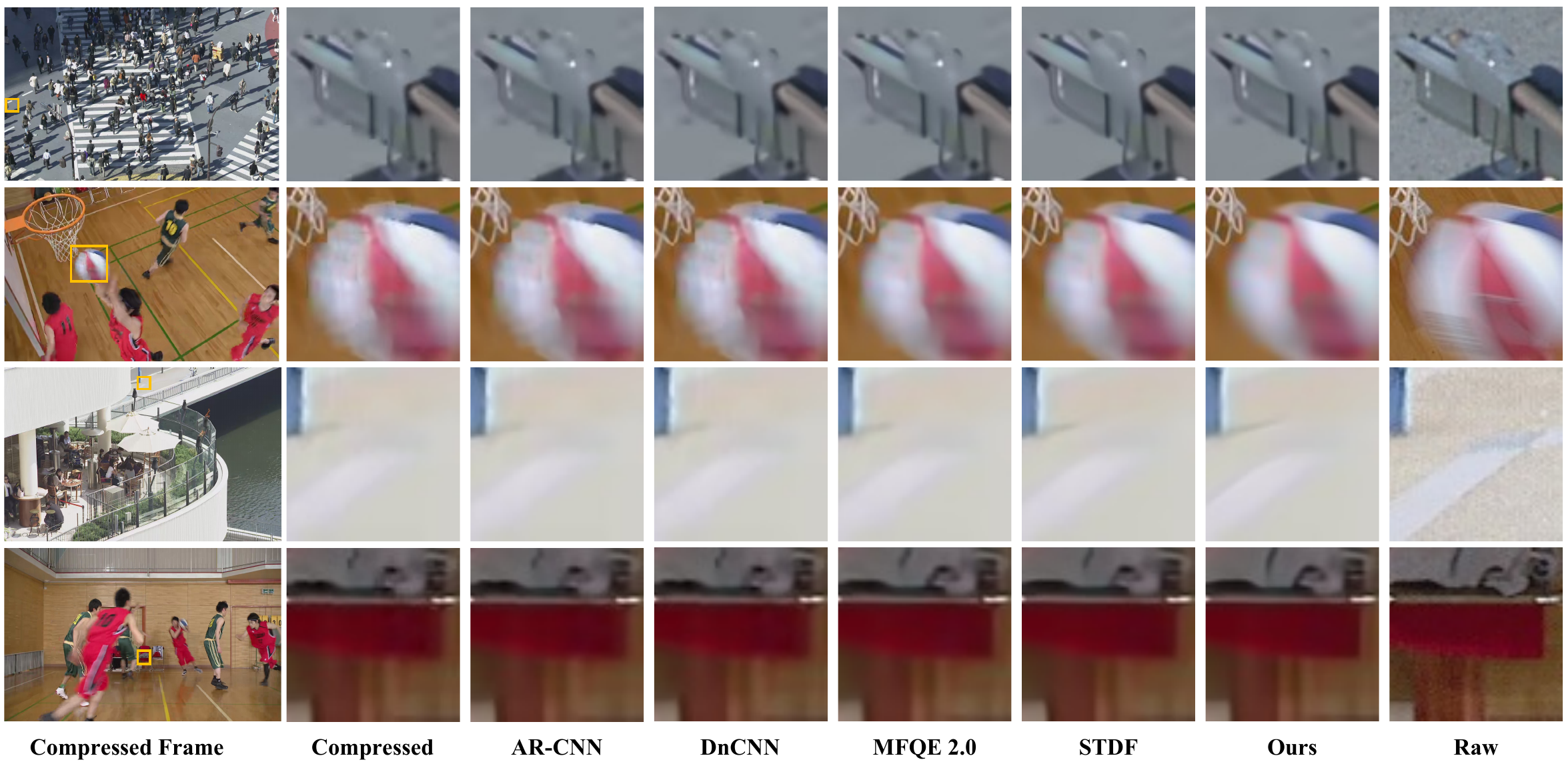}
	
	\caption{Qualitative examples at $QP$=37.}
	\label{fig-case}
\end{figure*}

\begin{figure}
	\centering
	\includegraphics[width=0.9\linewidth]{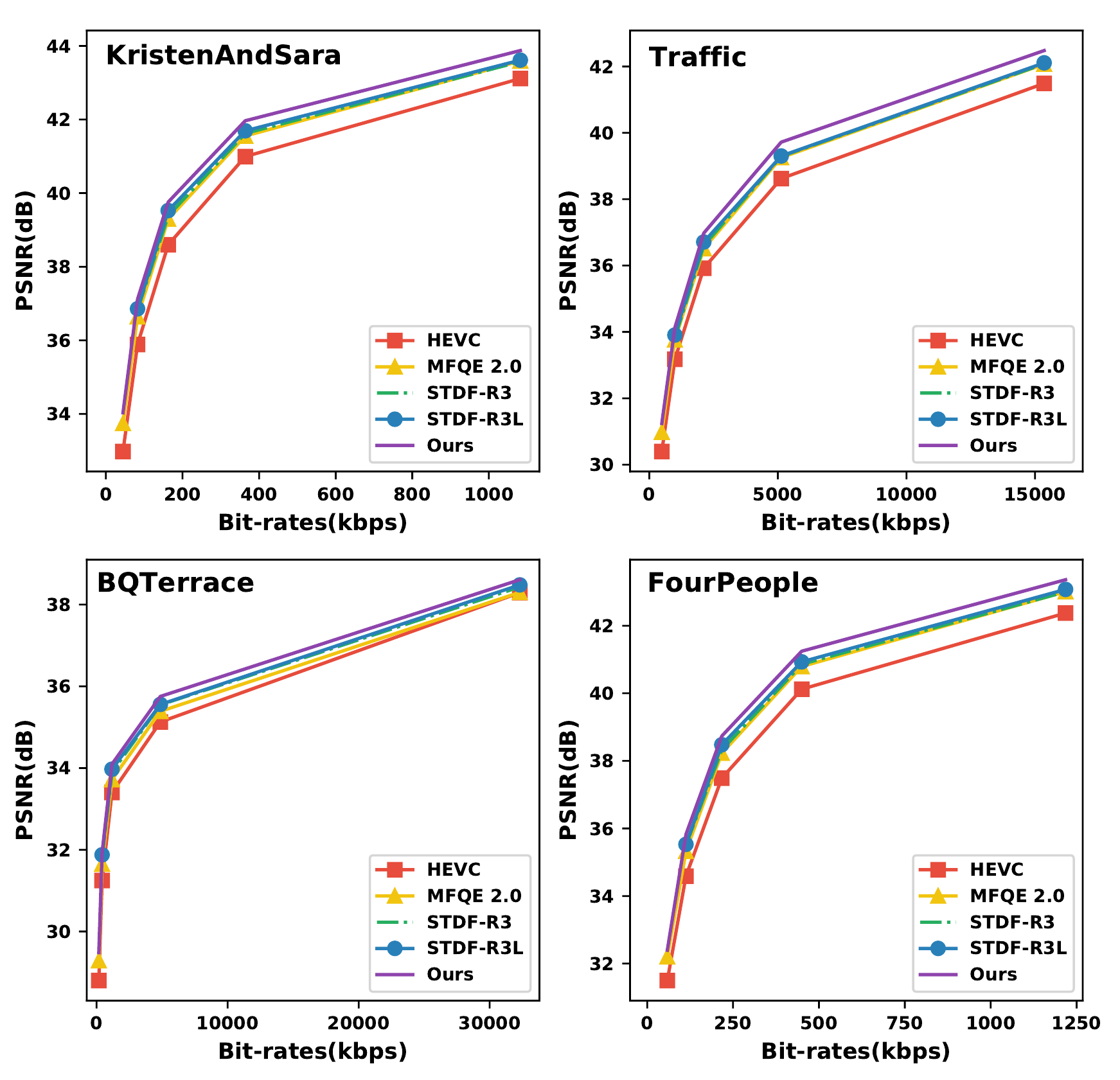}
	\caption{Rate-distortion performance of four test sequences.}
	\label{fig-rd}
\end{figure}

\textbf{Quality fluctuation.}
Quality fluctuation is another observable measurement for the overall quality of enhanced videos. Drastic quality fluctuation of frames accounts for severe texture shaking and degradation of the quality of experience~(QoE).  Therefore, we present two PSNR curves of test sequences compressed by HEVC, the corresponding sequences enhanced by STDF and our method in Fig.~\ref{fig-fluc}. Comparing with STDF, our method achieves larger PSNR improvement over the compressed frames, especially for non-PQFs, which means that the quality of frames enhanced by our method fluctuates less than that by STDF. Besides, we also evaluate the fluctuation by Standard Deviation~(SD), and Peak-Valley Difference~(PVD) of each test sequence as in \cite{xu2019non, guan2019mfqe, yang2018multi}. Results of PSNR are presented in Tab.~\ref{tab-SDPVD}, and our method still achieves impressive results of SD and PVD, which means that our method performs more stably than the other methods.

\textbf{Rate-distortion performance.}
We then evaluate the rate-distortion of our method and compare it with state-of-the-art methods. For ease of exposition, we only present the results of compressed videos, the enhanced results of two state-of-the-art methods~(MFQE 2.0 and STDF) and our method in Fig.~\ref{fig-rd}. Here, we do not show the results of STDF at $QP$=42 due to the lack of data in \cite{deng2020spatio}.
From the curves in Fig.~\ref{fig-rd}, we can see that our method performs better than the state-of-the-art approaches in rate-distortion performance. Following the experiments in \cite{guan2019mfqe}, we also evaluate the BD-bitrate (BD-BR) reduction in Tab.~\ref{tab-bdr}, which is calculated over five PSNR results at $QP$=22, 27, 32, 37 and 42, while the result of STDF is obtained with four QPs. 
Average results of BD-BR reduction for MFQE 2.0, STDF, \emph{same QE} and \emph{IQE} are 14.06\%, 20.79\%, 22.49\% and 25.86\%, respectively.
These results show the advantage of our techniques, and the methods with our techniques can achieve much better QoE under the same bit rate.

\begin{table*}
	\setlength{\linewidth}{\textwidth}
	\setlength{\hsize}{\textwidth}
	\centering
	\scriptsize
	\caption{Overall BD-BR reduction ($\%$) of test sequences with the HEVC baseline as an anchor. Calculated at QP = 22, 27, 32, 37 and 42.}
	\begin{tabular}{|c|l|| c@{\hspace{0.2em}} c @{\hspace{0.4em}}c @{\hspace{0.4em}}c @{\hspace{0.4em}}c@{\hspace{0.4em}} c@{\hspace{0.4em}} c @{\hspace{0.4em}}c@{\hspace{0.4em}} | c@{\hspace{0.4em}} c|}
		
		\hline
		\multicolumn{2}{|c||}{Sequence} & AR-CNN~\cite{dong2015compression} & DnCNN~\cite{zhang2017beyond} & Li \textit{et al.}~\cite{li2017efficient} & DCAD~\cite{wang2017novel} & DS-CNN~\cite{yang2017decoder} & MFQE 1.0~\cite{yang2018multi} & MFQE 2.0~\cite{guan2019mfqe} & STDF~\cite{deng2020spatio} & \emph{same QE} & \emph{IQE} \\
		
		\hline
		\multirow{2}{*}{A} & \textit{Traffic}
		& 7.40 & 8.54 & 10.08 & 9.97 & 9.18 & 14.56 & 16.98 & 21.19 & \textbf{22.11} & \textbf{26.99} \\
		
		& \textit{PeopleOnStreet}
		& 6.99 & 8.28 & 9.64 & 9.68 & 8.67 & 13.71 & 15.08 & 17.42 & \textbf{20.59} & \textbf{24.25} \\
		
		\cline{1-12}
		\multirow{5}{*}{B} & \textit{Kimono}
		& 6.07 & 7.33 & 8.51 & 8.44 & 7.81 & 12.60 & 13.34 & 17.96 & \textbf{22.73} & \textbf{25.01} \\
		
		& \textit{ParkScene}
		& 4.47 & 5.04 & 5.35 & 5.68 & 5.42 & 12.04 & 13.66 & 18.10 & \textbf{20.67} & \textbf{25.00} \\
		
		& \textit{Cactus}
		& 6.16 & 6.80 & 8.23 & 8.69 & 8.78 & 12.78 & 14.84 & 21.54 & \textbf{22.65} & \textbf{25.33} \\
		
		& \textit{BQTerrace}
		& 6.86 & 7.62 & 8.79 & 9.98 & 8.67 & 10.95 & 14.72 & 24.71 & \textbf{25.85} & \textbf{28.98}\\
		
		& \textit{BasketballDrive}
		& 5.83 & 7.33 & 8.61 & 8.94 & 7.89 & 10.54 & 11.85 & 16.75 & \textbf{21.35} & \textbf{23.14}\\
		
		\cline{1-12}
		\multirow{4}{*}{C} & \textit{RaceHorses}
		& 5.07 & 6.77 & 7.10 & 7.62 & 7.48 &8.83& 9.61 & 15.62 & \textbf{16.12} & \textbf{17.99}\\
		
		& \textit{BQMall}
		& 5.60 & 7.01 & 7.79 & 8.65 & 7.64 & 11.11 & 13.50 & 21.12 & \textbf{23.90} & \textbf{26.81}\\
		
		& \textit{PartyScene}
		& 1.88 & 4.02 & 3.78 & 4.88 & 4.08 & 6.67 & 11.28 & 22.24 & \textbf{22.81} & \textbf{26.92}\\
		
		& \textit{BasketballDrill}
		& 4.67 & 8.02 & 8.66 & 9.80 & 8.22 & 10.47 & 12.63 &  15.94 & \textbf{18.32} & \textbf{20.13}\\
		
		\cline{1-12}
		\multirow{4}{*}{D} & \textit{RaceHorses}
		& 5.61 & 7.22 & 7.68 & 8.16 & 7.35 & 10.41 & 11.55 & 15.26 & \textbf{18.76} & \textbf{20.93}\\
		
		& \textit{BQSquare}
		& 0.68 & 4.59 & 3.59 & 6.11 & 3.94 & 2.72 & 11.00 & 33.36 & 31.84 & \textbf{39.08}\\
		
		& \textit{BlowingBubbles}
		& 3.19 & 5.10 & 5.41 & 6.13 & 5.55 & 10.73 & 15.20 & 23.54 & \textbf{23.86} & \textbf{28.76}\\
		
		& \textit{BasketballPass}
		& 5.11 & 7.03 & 7.78 & 8.35 & 7.49 & 11.70 & 13.43 & 18.42 & \textbf{21.65} & \textbf{24.37}\\
		
		\cline{1-12}
		\multirow{3}{*}{E} & \textit{FourPeople}
		& 8.42 & 10.12 & 11.46 & 12.21 & 11.13 & 14.89 & 17.50 & 22.91 & 22.61 & \textbf{26.48}\\
		
		& \textit{Johnny}
		& 7.66 & 10.91 & 13.05 & 13.71 & 12.19 & 15.94 & 18.57 & 24.55 & \textbf{24.60} & \textbf{27.82}\\
		
		& \textit{KristenAndSara}
		& 8.94 & 10.65 & 12.04 & 12.93 & 11.49 & 15.06 & 18.34 & 23.64 & \textbf{24.40} & \textbf{27.51}\\
		
		\cline{1-12}
		\multicolumn{2}{|c||}{Average} & 5.59 & 7.36 & 8.20 & 8.89 & 7.85 & 11.41 & 14.06 & 20.79 & \textbf{22.49} & \textbf{25.86}\\
		\hline
		
	\end{tabular}
	\label{tab-bdr}
\end{table*}

\subsection{Qualitative Comparison}
We also conduct qualitative comparison and present several visual examples at $QP$=37 in Fig.~\ref{fig-case}.
We can see that the compressed frames suffer severe compression artifacts (\emph{e.g.} missing vertical stripes, blocking artifact on the basketball). For the existing methods from the third to sixth columns, the enhanced patches are distorted by over-smoothing and temporal noise. However, our method restores much more detail or texture than the other methods. Compared with the baseline STDF, our method restores more details, especially for high-frequency information, such as sharpening edges. This means that by applying the technique introduced in our paper, multi-frame approaches can do restoration better than the original ones.

\begin{figure*}
	\centering
	\includegraphics[width=0.92\linewidth]{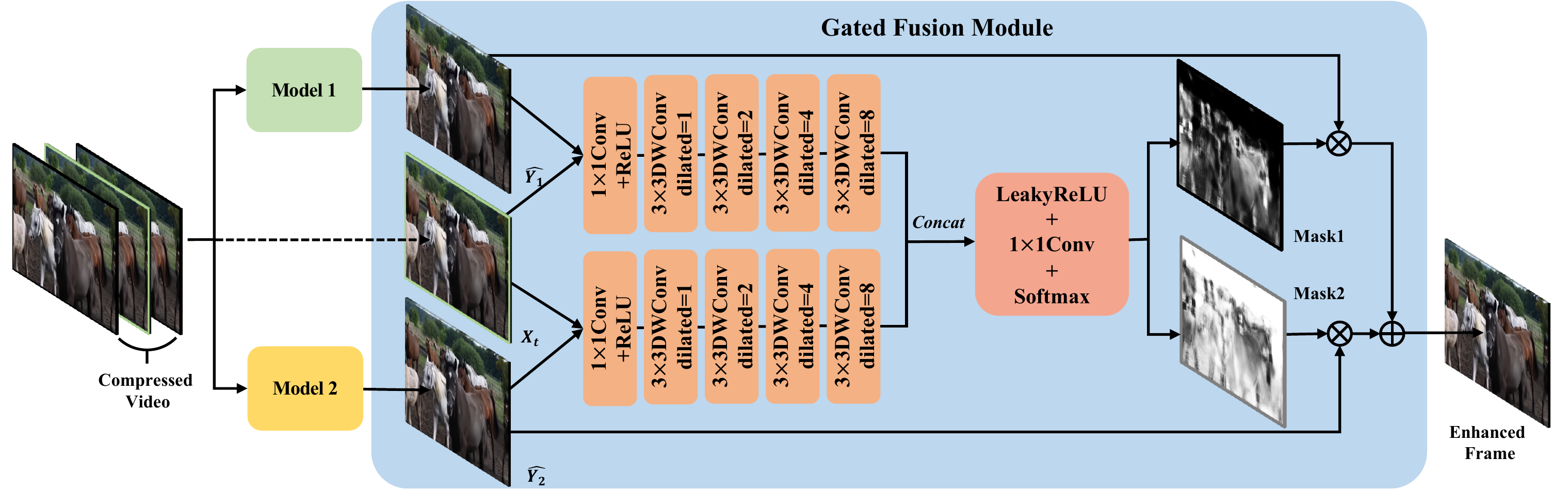}
	\caption{The architecture of gated fusion module.}
	\label{fig-gate}
\end{figure*}

\begin{table*}
	\caption{The final testing results of NTIRE Challenge on quality enhancement of heavily compressed videos.}
	\centering
	\scriptsize
	\begin{tabular}{c|cc|c|ccccc|c|cc}
		\hline
		\multicolumn{3}{c|}{Track 1} & \multicolumn{6}{c|}{Track 2} & \multicolumn{3}{c}{Track 3} \\
		\hline
		Ranking & PSNR(dB) & MS-SSIM & Ranking & MOS$\uparrow$ &LPIPS$\downarrow$ & FID$\downarrow$ & VID$\downarrow$ & VMAF$\uparrow$ & Ranking & PSNR~(dB) & MS-SSIM \\
		\hline
		1~(\textbf{Ours}) & \textbf{32.52}  & \textbf{0.9562} & 1~(\textbf{Ours}) & \textbf{71} & \textbf{0.0429} & \textbf{32.17} & \textbf{0.0137} & 75.69 & 1 & \textbf{30.37} & \textbf{0.9484}\\
		2 & 32.49 & 0.9552 & 2 & 69 & 0.0483 & 34.64 & 0.0179 & 71.55 & 2~(\textbf{Ours}) & 29.95 & 0.9468 \\
		3 & 32.04 & 0.9493 & 3 & 67 & 0.0561 & 46.39 & 0.0288 & 68.92 & 3 & 29.69 & 0.9423 \\
		4 & 31.90 & 0.9480 & 4 & 63 & 0.0561 &  50.61 & 0.0332 & 69.06 & 4 & 29.64 & 0.9405 \\
		5 & 31.86 & 0.9472 & 5 & 60 & 0.1018  &   72.27 & 0.0561 & \textbf{78.64} & 5 & 29.56 & 0.9403  \\
		\hline
	\end{tabular}
	\label{tab-NTIRE}
	
\end{table*}

\subsection{NTIRE 2021 Challenge}
In the NTIRE 2021 Challenge on Quality enhancement of compressed videos~\cite{yang2021ntire_method}, we won Track 1 and Track 2, and were the 2nd in Track 3. Detailed results are included in Tab.~\ref{tab-NTIRE}.
Besides the techniques introduced above, the performance also relies on a much deeper IQE module and two ensemble strategies, \emph{i.e.}, self-ensemble and gated fusion.

\textbf{Deeper IQE module.} In the competition implementation, we employ the IQE module with more Ada-WDSR-A-Blocks and wider features. Specifically, the number of channels for feature and blocks of Ada-Blocks in the deeper IQE module are 128 and 96, respectively. Thus, the number of feature channels in Ada-WDSR-A-Block is implemented as \{64, 256, 64\}.

\textbf{Self-ensemble.} In the competition, we utilize the self-ensemble strategy~\cite{ahn2018self} that can boost the restoration through multiple trails of inputs with different augmentation operations. Unlike the conventional ensemble strategies that integrate results from multiple models, self-ensemble takes the frames transformed by different augmentation operations, and averages these different but related outputs with the original to obtain the final predictions. In the competition, eight augmentation operations are exploited for evaluation.
Empirically, experimental results on the validation dataset in Track 1 show that STDF with basic IQE module~(shallow model) and deeper IQE module~(deep model) can achieve 0.2 and 0.12 dB PSNR improvement by utilizing self-ensemble.

\textbf{Gated Fusion module.}
Due to the limited official training data provided by the competition, models trained with only these data are easily dominated by the bias of training data. Meanwhile, limited clips mean rare scenes to be seen, but many unseen patterns may appear in inference, which restricts the performance of the model.
However, directly using large-scale data collected by ourselves will destroy the original distribution of training data. To minimize the offset between the two datasets and gain benefit from extra data, we propose a novel module to improve the performance of enhancement at the bottom of the pipeline. As illustrated in Fig.~\ref{fig-gate}, though each model has the same architecture~(STDF with deeper IQE), one is trained on the official training sets, and the other is on the extra videos crawled from Bilibili and YouTube, named as BiliTube4k. Inspired by \cite{tian2019versatile}, we exploit a stack of layers to output the mask and aggregate the predictions of two models via produced mask.
As shown in the middle of Fig.~\ref{fig-gate}, the mask $M$ in gated fusion module is of the same resolution of the target frame ranging from $[0,1]$. Thus, the output of gated fusion module can be
formulated as
$\hat{Y} = M \bigotimes \hat{Y}_1 \bigoplus (1-M)\bigotimes \hat{Y}_2$.
The detail of network architecture can be referred to Fig.~\ref{fig-gate}. Furthermore, such a structure of gated fusion module can be used in more models.

\textbf{Other details.} In Track 3, we utilize the model pre-trained in Track 1 as the pre-trained model, and then fine-tune it on training data of Track 3 with early stopping.
As for Track 2, we reuse and freeze the models from Track 1, and attach ESRGAN~\cite{wang2018esrgan} at the bottom of them. Specifically, we use the ESRGAN pre-trained on DIV2K dataset~\cite{agustsson2017ntire}, remove the pixel shuffle layer, and employ the FFT loss. Then, two ESRGANs trained on different datasets are integrated with the gated fusion module to produce the final enhanced frames.

\section{Conclusion}
In this paper, we present a method for improving existing multi-frame approaches in video compression artifact reduction via integrating multiple frames and frequency domain information.
Our method was developed for the NTIRE 2021 Challenge on Quality enhancement of heavily compressed videos Challenge, and won Track 1 and Track 2, and the 2nd place in Track 3. Through extensive experiments, we show that both our proposed reference frame proposal strategy and the FFT loss can achieve superior performance over state-of-the-art methods. In the future, more verification of our techniques is expected to be conducted on other multi-frame approaches.

\section{Acknowledgement}
Yi Xu, Minyi Zhao and Shuigeng Zhou were
supported by 2019 Special Fund for Artificial Intelligence
Innovation \& Development, Shanghai Economy and Information
Technology Commission (SHEITC), and partially by Science and
Technology Commission of Shanghai Municipality Project
(No.~19511120700).

{\small
\bibliographystyle{ieee_fullname}
\bibliography{egbib}
}

\end{document}